\documentclass[12pt]{article}
\usepackage{amsmath}

\renewcommand{\theequation}{\thesection.\arabic{equation}}
\csname @addtoreset\endcsname{equation}{section}
\newcommand{\eq}{\begin{equation}}
\newcommand{\feq}{\end{equation}}
\newcommand{\eqn}{\begin{eqnarray}}
\newcommand{\feqn}{\end{eqnarray}}
\newcommand{\arr}{\begin{eqnarray*}}
\newcommand{\farr}{\end{eqnarray*}}

\begin{document}

\begin{titlepage}
\begin{flushright}
HUTP-01/A017\\
CAMS/01-03\\
hep-th/0104044
\end{flushright}
\vspace{.3cm}
\begin{center}
\renewcommand{\thefootnote}{\fnsymbol{footnote}}
{\Large A Supersymmetric Solution in $N=2$ Gauged Supergravity with the Universal  Hypermultiplet}
\vfill
{\large \bf        Michael Gutperle$^1$\footnote{email:
    gutperle@democritus.harvard.edu  }
and Wafic A.  Sabra$^2$\footnote{email: ws00@aub.edu.lb}}

\renewcommand{\thefootnote}{\arabic{footnote}}
\setcounter{footnote}{0}
\vfill
{\small
$^1$Jefferson Physical Laboratory, Harvard University\\
Cambridge, MA 02138, USA\\
\vspace*{0.4cm}
$^2$ Center for Advanced Mathematical Sciences (CAMS)
and\\
Physics Department, American University of Beirut, Lebanon.\\}
\end{center}
\vfill
\begin{center}
{\bf Abstract}
\end{center}
 We present  supersymmetric  solutions for the theory
of
gauged supergravity in five dimensions
obtained by  gauging the shift symmetry of the axion of the
universal hypermultiplet.
This gauged theory  can also  be obtained by dimensionally
reducing M-theory on a Calabi-Yau threefold with background flux. The
solution found  preserves half
of the $N=2$ supersymmetry, carries electric fields and has
nontrivial scalar field representing  the CY-volume.
 We comment on the possible solutions of more general
hypermultiplet
gauging.

\end{titlepage}

\section{Introduction}

The study of supersymmetric configurations in five-dimensional gauged
supergravity theories has been receiving considerable attention in recent
years. This study is relevant to AdS/CFT correspondence which relates string
theory (or, for energies much smaller than the string scale, supergravity)
on anti-de~Sitter (AdS) spaces to supersymmetric conformal field theories
residing on the boundary of AdS \cite{adscft}. This provides a possibility
to \ study the regime of large $N$ and large t'Hooft coupling of these field
theories by means of classical supergravity solutions. It is hoped that this
correspondence will eventually lead to a deeper insight into confinement in
QCD, or at least into the strong coupling regime of gauge theories with
fewer supersymmetries. Moreover, gauged supergravity models in five
dimensions provide the framework which may incorporate smooth
Randall-Sundrum brane worlds \cite{RS}. Most of the analysis so far has been
focused on the study supersymmetric backgrounds in the gauged supergravity
with vector multiplet gaugings \cite{dr, bh}. In these models, the so-called 
$U(1)$-gauged theories, the hypermultiplets decouple in the analysis of
finding supersymmetric background solutions. Apart from the
hyperbolic-string solutions \cite{bh1}, most of the known supersymmetric
solutions contain naked singularities or have other peculiar properties. For
example, the rotating charged solutions in the five dimensional gauged
theories were found to violate causality and represent naked time machines 
\cite{bh2}. Moreover, domain wall solutions have also been the subject of
intense research activity motivated both by the interpretation of domain
wall solutions as realizations of renormalization group flows in the dual
field theory and by the attempt to find an embedding of Randall-Sundrum
scenario in string or M-theory (see \cite{KL} and references therein).

Five dimensional supergravity theory needs a minimum of eight supercharges.
The coupling of this theory to vector multiplets and its gaugings was
formulated in \cite{gst1, gst2}. The gravity multiplet contains the
graviton, the gravitino and the graviphoton. The abelian vector multiplets
each contains a $U(1)$ vector field, a gaugino and a real scalar field. The
scalar fields of the vector multiplets parameterize a space $M_{V}$ defining
a very special geometry where all the couplings of the theory are determined
in terms of a cubic form $\mathcal{F}=C_{IJK}h^{I}h^{J}h^{K}$ \cite{dWvP}.

The compactification of M-theory on Calabi-Yau three-fold \cite{cyc} gives
an effective field theory which is $N=2$ five dimensional supergravity
interacting with a number of hypermultiplets and vector multiplets with
scalars parameterizing a manifold $\mathcal{M=}M_{V}\times $ $M_{H}.$ In this
framework, $h^{I}$ are related to the K\"{a}hler class moduli fields and $%
C_{IJK}$ are the topological intersection numbers. The hypermultiplets, each
containing two hyperinos and four real scalar fields living on a
quaternionic K\"{a}hler manifold $M_{H}\cite{sierra, Bagger}.$ In string or
M-theory compactification, there is at least one hypermultiplet which
contains the Calabi-Yau volume. The geometry of this hypermultiplet is given
by the coset space $\frac{SU(2,1)}{U(2)}$ \cite{Ferrara}. Gauged
supergravity theories can be obtained by gauging the isometries of the
manifolds $M_{V}$ and $M_{H}$ \cite{Andrianopoli} (see also \cite{ceresole}
and references therein)$.$ If one compactifies M-theory on a Calabi-Yau
threefold in the presence of $G$-flux, the effective theory will contain a
superpotential which is related to the gauging of some global isometries of
the scalar manifold \cite{lukas, bg}. Due to the non-trivial flux and the
Chern-Simons term in the eleven dimensional supergravity, the axionic scalar
of the hypermultiplet becomes charged and the theory obtained corresponds to
gauging a combination of the R-symmetry and the axionic shift symmetry.

\bigskip

In general BPS solutions are characterized by their symmetries. Flat BPS
domain walls have an $ISO(3,1)$ Poincare symmetry and the solution depends
only on the transverse coordinate \footnote{
 Curved domain walls in four dimensional $N=2$ gauged supergravity were
recently discussed in \cite{lustb}}. In this paper,  we will consider
solutions which have $SO(4)$ rotational symmetry and where the metric
depends only on the radial coordinate. Such supersymmetric solutions with
hypermultiplet gauging  have not been yet explored. In these cases, the
hypermultiplets do not decouple from the theory and this seems to rule out
the presence of BPS solutions, at least when the gauge fields are set to
zero \cite{ceresole}. Therefore, it seems that in order to obtain
supersymmetric solutions, one needs to have non--trivial electric or
magnetic fields. It is our purpose in this paper to touch upon this subject
by considering the case where the axionic shift of the hypermultiplet is
also gauged. This model comes from the dimensional reduction of M-theory on
a Calabi-Yau manifold with background flux. We organize our work as follows.
In section two, we review the gauged supergravity model we wish to study,
the reader is referred to \cite{ceresole} for a more detailed discussion. In
section three we present a supersymmetric solution by examining the
supersymmetry transformations of the fermionic fields as well as the
equations of motion. Section four contains a summary and a discussion.

\section{The Gauged Supergravity Model}

In the following, we will discuss the five dimensional gauged supergravity
theory with one hypermultiplet only. A hypermultiplet is always present in
any Calabi-Yau (CY) compactification of M-theory and type II string theory.
For example, compactifying M-theory or type IIA string theory on a rigid CY
( i. e. $h_{2,1}=0$) leads to an $N=2$ theory with a single hypermultiplet,
the so-called universal hypermultiplet \cite{cfg} (the term ``universal'' is
slightly misleading for compactifications with $h_{2,1}>0$, see \cite
{Aspinwall} section 4.4.3). The bosonic part of the action for the universal
hypermultiplet which can be derived by dimensionally reducing eleven
dimensional supergravity is given by 
\begin{eqnarray}
S_{hyper} &=&-\int d^{5}x\sqrt{h}\Big((\partial _{\mu }\phi )^{2}+e^{2\phi }%
\big((\partial _{\mu }\chi _{1})^{2}+(\partial _{\mu }\chi _{2})^{2}\big) 
\notag  \label{eq:hypact} \\
&&+{\frac{1}{4}}e^{4\phi }\big(\partial _{\mu }a+(\chi _{1}\partial _{\mu
}\chi _{2}-\chi _{2}\partial _{\mu }\chi _{1})\big)^{2}\Big).
\end{eqnarray}

Classically$\footnote{%
The moduli space receives gravitational correction which depends on the
Euler character of the Calabi-Yau threefold,\textbf{\ }this can however be
absorbed by a field redefinition \cite{strominger}.}$, the scalar fields of
this action parameterize the group coset manifold $SU(2,1)/U(2).$ This can be
easily seen if one defines the new complex coordinates\footnote{%
In the above action, $\phi $ is associated with the volume of the CY, $a$
comes from the dual of the four-form of eleven dimensional supergravity, $%
F=dA_{3}$, and $C$ corresponds to the expectation values of $A_{3}.$} 
\begin{equation}
S=e^{-2\phi }+ia+\chi _{1}^{2}+\chi _{2}^{2},\quad C=\chi _{1}+i\chi _{2}.
\end{equation}
The moduli space is K\"{a}hler with K\"{a}hler potential 
\begin{equation}
\mathcal{K}=\phi =-{\frac{1}{2}}\ln \Big({\frac{S+\bar{S}}{2}}-|C|^{2}\Big).
\label{eq:kaehlp}
\end{equation}
In the appendix, we gather all the necessary formulae of the quaternionic
geometry defined by the universal hypermultiplet as well as the various
quantities resulting from gauging the axionic shift symmetry.

In addition to the hypermultiplet action, we also have the $N=2$
supergravity multiplet whose bosonic action is given by 
\begin{equation}
S_{grav}=-\int d^{5}x\sqrt{h}\Big({\frac{1}{2}}R+{\frac{1}{4}}F_{\mu \nu
}F^{\mu \nu }\Big)+{\frac{1}{6\sqrt{6}}}\int d^{5}x\epsilon ^{\mu \nu \rho
\sigma \tau }F_{\mu \nu }F_{\rho \sigma }A_{\tau }.  \label{ggaction}
\end{equation}
The supersymmetry transformations of the ungauged theory for the fermions
are given by \cite{sierra,Bagger} 
\begin{eqnarray}
\delta \psi _{\mu i} &=&\partial _{\mu }\epsilon _{i}+{\textstyle{\frac{1}{4}%
}}\omega _{\mu }^{ab}\gamma _{ab}\epsilon _{i}+\partial _{\mu
}q^{u}(p_{i}^{j})_{u}\epsilon _{j}+{\frac{i}{4\sqrt{6}}}(\gamma _{\mu \nu
\rho }-4\gamma _{\mu \nu }\gamma _{\rho })\epsilon _{i}F^{\nu \rho },  \notag
\\
\delta \zeta ^{A} &=&-{\frac{i}{2}}f_{iu}^{A}\gamma ^{\mu }\epsilon
^{i}\partial _{\mu }q^{u}.
\end{eqnarray}
The gauged $N=2$ supergravity theory can be obtained by gauging the $R$%
-symmetry as well as isometries in the hypermultiplet manifold. The isometry
in the hypermultiplet moduli space is associated with a Killing vector $k^{u}
$ 
\begin{equation}
q^{u}\rightarrow q^{u}+\varepsilon k^{u}(q).  \label{eq:isoma}
\end{equation}
The $N=2$ supersymmetry demands that $k^{u}$ is determined in terms of a
triplet of Killing prepotentials $P_{j}^{i}$ 
\begin{equation}
k^{u}(K_{uv})_{j}^{i}=\partial _{v}P_{j}^{i}+[p_{v},P]_{j}^{i},
\label{eq:prepotn}
\end{equation}
where $(K_{uv})_{i}^{j}$ is the $SU(2)$ triplet of K\"{a}hler forms and $%
 p_{i}^{j}$ is the $SU(2)$ connection. In order to restore
supersymmetry in the gauged theory, the bosonic part of the action \footnote{%
Since we are interested in bosonic BPS solutions, we will neglect fermionic
mass terms induced by the gauging.} of the ungauged theory gets modified in
two ways: Firstly the derivatives and connections get replaced by gauge
covariant ones and secondly a potential term gets added to the action $%
S_{pot}=-\int d^{5}x\sqrt{-h}V$ where 
\begin{equation}
V(q)=-g^{2}P_{ij}P^{ij}+2g^{2}N_{iA}N^{iA}.
\end{equation}
The supersymmetry transformations also get modified \cite{ceresole} 
\begin{eqnarray}
\delta \psi _{\mu i} &=&\mathcal{D}_{\mu }\epsilon _{i}+{\frac{i}{4\sqrt{6}}}%
(\gamma _{\mu \nu \rho }-4\gamma _{\mu \nu }\gamma _{\rho })\epsilon
_{i}F^{\nu \rho }+{\frac{i}{\sqrt{6}}}g\gamma _{\mu }\epsilon ^{j}P_{ij}, 
\notag \\
\delta \zeta ^{A} &=&-{\frac{i}{2}}f_{iu}^{A}\gamma ^{\mu }\epsilon
^{i}D_{\mu }q^{u}+g\epsilon ^{i}N_{i}^{A}.
\end{eqnarray}
Where 
\begin{equation}
D_{\mu }\epsilon _{i}=\partial _{\mu }\epsilon _{i}+{\textstyle{\frac{1}{4}}}%
\omega _{\mu }^{ab}\gamma _{ab}\epsilon _{i}+\partial _{\mu
}q^{u}(p_{i}^{j})_{u}\epsilon _{j}+gk^{u}A_{\mu }(p_{i}^{j})_{u}\epsilon
_{j}+gA_{\mu }P_{i}^{j}\epsilon _{j}.  \label{covdgauged}
\end{equation}

There are many different isometries which can be gauged for the universal
hypermultiplet \footnote{%
See \cite{BC} for an exhaustive treatment of the possible gaugings. Note
that for the translational gauging considered in our paper the last two
terms in (\ref{covdgauged}) cancel.}. Here we will consider the gauging of
the shift symmetry of the axion $a\rightarrow a+const$. This gauging has
special significance since, as mentioned in the introduction, corresponds to
turning on fluxes on the internal Calabi-Yau threefold \cite{lukas, bg} .

\section{The Supersymmetry Preserving  Solution}

In this section we present a field configuration which preserves half of the 
$N=2$ supersymmetries. Domain wall solutions of this model have been
discussed in \cite{lukas, bg}. As an ansatz, we will only keep the component 
$A_{t}$ of the graviphoton potential and the volume scalar $\phi $ as the
dynamic fields. We should point out that our analysis is equally valid for
the general gauged supergravity models with vector multiplets with constant
scalar fields. The most general ansatz for the metric which has $SO(4)$
symmetry in spherical coordinates is given by 
\begin{equation}
ds^{2}=-e^{2V}(dt)^{2}+e^{2W}\left( dr^{2}+f^{2}r^{2}(d\theta ^{2}+\sin
^{2}\theta d\phi ^{2}+\cos ^{2}\theta d\psi ^{2})\right) .  \label{metric}
\end{equation}

The F\"{u}nfbein and its inverse can be chosen as 
\begin{eqnarray}
e_{t}^{0} &=&e^{V},\qquad e_{r}^{1}=e^{W},\qquad e_{\theta
}^{2}=e^{W}fr,\qquad e_{\phi }^{3}=e^{W}fr\sin \theta ,\qquad e_{\psi
}^{4}=e^{W}fr\cos \theta ,  \notag \\
e_{0}^{t} &=&e^{-V},\text{ \ \ \ \ \ }e_{1}^{r}=e^{-W},\qquad e_{2}^{\theta
}=\frac{e^{-W}}{fr},\text{ \ \ \ }e_{3}^{\phi }=\frac{e^{-W}}{fr\sin \theta }%
,\qquad e_{4}^{\psi }=\frac{e^{-W}}{fr\cos \theta }.
\end{eqnarray}

$\allowbreak $The non-vanishing components of the spin connection are given
by 
\begin{eqnarray}
\omega _{t}^{01} &=&\partial _{r}Ve^{V-W},  \notag \\
\omega _{\theta }^{12} &=&-(fr\partial _{r}W+r\partial _{r}f+f),  \notag \\
\omega _{\phi }^{13} &=&-(fr\partial _{r}W+r\partial _{r}f+f)\sin \theta , 
\notag \\
\omega _{\psi }^{14} &=&-(fr\partial _{r}W+r\partial _{r}f+f)\cos \theta , 
\notag \\
\omega _{\phi }^{23} &=&-\cos \theta ,  \notag \\
\omega _{\psi }^{24} &=&\sin \theta .
\end{eqnarray}

As a first step, we consider the supersymmetry transformations of the
hyperinos. Allowing only a non-trivial $\phi $, these transformations for
our ansatz give, 
\begin{eqnarray}
\delta \zeta ^{1} &=&-{\frac{i}{4}}e^{-W}e^{2\phi }\partial _{r}(e^{-2\phi
})\gamma _{1}\epsilon ^{2}-{\frac{1}{4}g}e^{-V}e^{2\phi }A_{t}\gamma
_{0}\epsilon ^{2}+i{\frac{\sqrt{6}}{8}}ge^{2\phi }\epsilon ^{2},  \notag \\
\delta \zeta ^{2} &=&-{\frac{i}{4}}e^{-W}e^{2\phi }\partial _{r}(e^{-2\phi
})\gamma _{1}\epsilon ^{1}+{\frac{1}{4}g}e^{-V}e^{2\phi }A_{t}\gamma
_{0}\epsilon ^{1}-i{\frac{\sqrt{6}}{8}}ge^{2\phi }\epsilon ^{1}.
\end{eqnarray}
Here and in the following all functions will only depend on the radial
parameter $r$ and derivatives with respect to $r$ will be denoted by primes.
The hyperino transformations vanish for half the supersymmetries if $%
\epsilon ^{1,2}$ $\mathbf{\ }$satisfy

\begin{equation}
\left( ia\gamma _{0}+b\gamma _{1}\right) \epsilon ^{1}=\epsilon ^{1},\quad
\quad \left( ia\gamma _{0}-b\gamma _{1}\right) \epsilon ^{2}=\epsilon ^{2}.
\label{proje}
\end{equation}

\bigskip The vanishing of the hyperinos supersymmetry transformations thus
gives 
\begin{eqnarray}
\epsilon ^{2} &=&\sqrt{\frac{2}{3}}\left( e^{-W}(e^{-2\phi })^{\prime }\frac{%
1}{g}\gamma _{1}-ie^{-V}A_{t}\gamma _{0}\right) \epsilon ^{2},  \notag \\
\epsilon ^{1} &=&\sqrt{\frac{2}{3}}\left( -{\frac{1}{g}}e^{-W}(e^{-2\phi
})^{\prime }\gamma _{1}-{i}e^{-V}A_{t}\gamma _{0}\right) \epsilon ^{1}
\end{eqnarray}

\bigskip

so $a$ and $b$ are given by 
\begin{equation}
a=-\sqrt{\frac{2}{3}}e^{-V}A_{t},\quad \quad b=-{\frac{1}{g}}\sqrt{\frac{2}{3%
}}e^{-W}(e^{-2\phi })^{\prime }.
\end{equation}
The condition for (\ref{proje}) to be projectors is $a^{2}+b^{2}=1$. The
time component of the gravitino supersymmetry transformation is given by

\begin{equation}
\delta \psi _{t}^{1}=\left[ \mathcal{\partial }_{t}+\frac{1}{2}V^{\prime
}e^{V-W}\gamma _{0}\gamma _{1}+\frac{i}{\sqrt{6}}e^{-W}A_{t}^{\prime }\gamma
_{1}+{\frac{1}{4\sqrt{6}}}ge^{2\phi }e^{V}\gamma _{0}\right] \epsilon ^{1}.
\label{gst}
\end{equation}
Here and in the following we will only display the $\delta \psi ^{1}$
variations and drop the spinor index, the $\delta \psi ^{2}$ work in the
same way after adjusting some signs. Assuming that $\partial _{t}\epsilon =0$%
, the projection gives the following condition

\begin{equation}
A_{t}=c_{1}e^{V}.
\end{equation}
This implies that $a=-\sqrt{\frac{2}{3}}c_{1}$ is a constant, and therefore $%
b$ is a constant too and one finds that 
\begin{equation}
(e^{-2\phi })^{\prime }=c_{2}ge^{W},
\end{equation}
where $b=-\sqrt{\frac{2}{3}}c_{2}$. The second condition coming from $\delta
\psi _{t}$ can then be brought into the following form 
\begin{equation}
V^{\prime }=-{\frac{g}{2\sqrt{6}b}}e^{2\phi }e^{W}.
\end{equation}
Using these relations the $r$ component of the $\delta \psi _{\mu }$ can be
simplified and gives 
\begin{equation}
\delta \psi _{r}=\left( \mathcal{\partial }_{r}-{\frac{1}{2}}V^{\prime
}\right) \epsilon ,
\end{equation}
which can easily be integrated. The other components of the gravitino
supersymmetry variations are 
\begin{eqnarray}
\left[ \mathcal{\partial }_{\theta }-\frac{1}{2}e^{-W}(e^{W}fr)^{\prime
}\gamma _{12}-\frac{i}{2\sqrt{6}}fre^{-V}A_{t}^{\prime }\gamma _{210}+\frac{1%
}{4\sqrt{6}}ge^{2\phi }e^{W}fr\gamma _{2}\right] \epsilon &=&0,  \notag \\
\left[ \mathcal{\partial }_{\phi }-\frac{1}{2}\sin \theta \Big(%
e^{-W}(e^{W}fr)^{\prime }\gamma _{13}+\frac{ifr}{\sqrt{6}}%
e^{-V}A_{t}^{\prime }\gamma _{103}-\frac{g}{2\sqrt{6}}e^{2\phi
}e^{W}fr\gamma _{3}\Big)-{\frac{1}{2}}\cos \theta \gamma _{23}\right]
\epsilon &=&0,  \notag \\
\left[ \mathcal{\partial }_{\psi }-\frac{1}{2}\cos \theta \Big(%
e^{-W}(e^{W}fr)^{\prime }\gamma _{14}+\frac{ifr}{\sqrt{6}}%
e^{-V}A_{t}^{\prime }\gamma _{104}-\frac{g}{2\sqrt{6}}e^{2\phi
}e^{W}fr\gamma _{4}\Big)+{\frac{1}{2}}\sin \theta \gamma _{24}\right]
\epsilon &=&0.  \notag \\
\end{eqnarray}
Integrability conditions $[\nabla _{\mu },\nabla _{\nu }]\epsilon =0,$ fix
all other functional relations and constants. One gets

\begin{equation}
fre^{W}e^{2\phi }=k,
\end{equation}
where $k$ is a constant. One also obtains the relation 
\begin{equation}
b=-\frac{1}{2\sqrt{6}c_{2}},
\end{equation}
which implies that the dilaton $\phi $ and metric function $V$ are related
by 
\begin{equation}
V=-2\phi .
\end{equation}
All the constants are thus fixed and given by 
\begin{equation}
c_{2}^{2}=\frac{1}{4},\text{ \ \ \ \ }c_{1}^{2}=\frac{5}{4},\text{ \ \ \ \ \ 
}a^{2}=\frac{5}{6},\text{ \ \ \ }b^{2}=\frac{1}{6},\text{ \ \ \ \ }%
k^{2}g^{2}=\frac{32}{15}.  \label{numbers}
\end{equation}

\bigskip

So far we have determined a configuration which preserves half of
supersymmetry. The gauge field $A_{t}$ and the scalar $\phi $ depend on the
metric which can be expressed in terms of the function $V,$ these relations
are summarized by

\begin{eqnarray}
A_{t} &=&c_{1}e^{V},  \notag \\
(e^{-2\phi })^{\prime } &=&(e^{V})^{\prime }=c_{2}ge^{W},  \notag \\
fre^{W} &=&ke^{V}.
\end{eqnarray}

In order to fix the space-time dependence of the configuration, one needs to
solve the equations of motion and determine the function $V.$

Einstein's equation reads,

\begin{equation}
G_{\mu \nu }=R_{\mu \nu }-{\frac{1}{2}}h_{\mu \nu }R=T_{\mu \nu },
\end{equation}

where, the stress-energy tensor is given by

\begin{equation}
T_{\mu \nu }={\frac{2}{\sqrt{-h}}}{\frac{\delta S}{\delta h^{\mu \nu }}}%
=g_{uv}D_{\mu }q^{u}D_{\nu }q^{v}-{\frac{1}{2}}h_{\mu \nu }g_{uv}D_{\rho
}q^{u}D^{\rho }q^{v}+F_{\mu \nu }^{2}-{\frac{1}{4}}h_{\mu \nu }F^{2}-h_{\mu
\nu }V.
\end{equation}

Since we look at a purely electric configuration the Chern-Simons term in the $N=2$ supergravity Lagrangian is not important and the gauge field and the scalar field equations of motion, are respectively
given by (we set $a=C=0)$

\begin{eqnarray}
\frac{1}{\sqrt{-h}}\partial _{\mu }\left( \sqrt{-h}F^{\mu \nu }\right) -{%
\frac{g^{2}}{2}}e^{4\phi }A^{\nu } &=&0, \\
\frac{1}{\sqrt{-h}}\partial _{\mu }\left( \sqrt{-h}h^{\mu \nu }\partial
_{\nu }\phi \right) -\frac{1}{2}g^{2}e^{4\phi }A_{\mu }A^{\mu }-{\frac{1}{4}}%
g^{2}e^{4\phi } &=&0.
\end{eqnarray}

It is customary that the supersymmetric property of the solution determines
the relations among the various fields of the configuration, i.e., an
off-shell solution which can only be fixed by solving the equations of
motion. For example, in the ungauged and the $U(1)$-gauged theory, solving
the gauge field equations fixes the solution in terms of harmonic functions 
\cite{bls, bh}. The harmonic functions are clearly solutions of a second
order differential equations and can not be determined from the first order
differential equations resulting from the vanishing of the supersymmetry
transformations \cite{bls}.

\bigskip

Here due to the second term in the gauge field equation of motion, it is
clear that our solution can not be determined in terms of a harmonic
function. As a matter of fact, it can be checked that all the above
equations of motion are satisfied for any choice of the function $e^{V}$.
This might come as a surprise. However, the BPS configuration we found is of
the following form 
\begin{equation}
ds^{2}=-e^{2V}dt^{2}+{\frac{4}{g^{2}}(}\partial
_{r}e^{V})^{2}dr^{2}+e^{2V}k^{2}dS_{3}^{2},  \label{solmet}
\end{equation}
\begin{equation}
A_{t}=c_{1}e^{V},\quad \quad e^{-2\phi }=e^{V}.
\end{equation}
By inspecting the metric (\ref{solmet}) it is clear why the equations of
motion did not determine $V$. This is because different $V$ correspond to
different choice of radial coordinates. Indeed defining a new coordinate $%
u(r)=e^{V}$ , we find that the solution in the new coordinates is given by 
\begin{equation}
ds^{2}=-u^{2}dt^{2}+{\frac{4}{g^{2}}}du^{2}+k^{2}u^{2}d\Omega _{S_{3}}^{2}
\end{equation}
and 
\begin{equation}
A_{t}=c_{1}u,\quad \quad e^{-2\phi }=u.
\end{equation}
The solution has several peculiar properties. For example, the fieldstrength 
$F_{ut}=c_{1}$ is constant and the charge is given by 
\begin{equation}
q=\oint \ast F=c_{1}c_{2}gk^{3}u^{2},
\end{equation}
diverges as $u\rightarrow \infty $. Since $e^{2\phi }$ parameterizes the
volume of the Calabi-Yau manifold, we see that the volume becomes infinite
at  $u=0$ and goes to zero at $u\rightarrow \infty $. The Ricci scalar is
given by $R=-{\frac{3}{8}\frac{1}{u^{2}}}$ and the metric is singular at $%
u=0.$  Note that for the values of the constants given in (\ref{numbers}), 
the spatial part of the metric has the form of a cone. It is an open
question whether the singular BPS solution we found has a proper
interpretation in M-theory, we note however that the solution shares some of
its properties with other BPS solutions in gauged supergravities.

As it was mentioned before, our solution is also valid in the presence of
vector multiplets with trivial scalars. In this case the gauge fields are
given by

\begin{equation}
A_{t}^{I}=c_{1}h^{I}e^{V},
\end{equation}
and we set $\alpha _{I}h^{I}=1,$ where $\alpha _{I}$ is the flux vector \cite
{bg}. (For this ansatz, the gaugino supersymmetry variation vanishes).
Generalization of our results to four dimensions is straightforward since
the hypermultiplet sector in four dimensions is the same as in five
dimensions. In four dimensions, we have $\phi =-\phi _{4}$ where $\phi _{4}$
is the four dimensional dilaton.

\section{Discussion}

In this paper we have studied supersymmetric configurations for \ a gauged
supergravity theory coming from the compactification of M-theory on a
Calabi-Yau threefold with a background flux. This effective gauged theory
correspond to a supergravity theory where the axionic shift of the universal
hypermultiplet is also gauged. The peculiar configuration found is perhaps
related to the ``run-away'' nature of the scalar potential. The domain wall
solution for the gauging considered in our paper was discussed in \cite{bg}
and has a similar singular behavior. However, one would hope to get smooth
and better behaved black hole solutions for more general gauging and
potentials with some fixed points. It is thus of interest to study more
general gaugings like those discussed in \cite{thomas, BC}. Also one should
be less restrictive and study configurations with more non-trivial scalar
fields of the universal hypermultiplet and vector multiplets. In general,
the connection piece in $D_{\mu }$,\textbf{\ }as well as $P_{ij}$, $%
f_{iu}^{A}$ and $N_{i}^{A}$ in the hyperino variations, which were our
starting point to determine the projection condition on the Killing spinors
all define $2\times 2$ matrices. Therefore, for the general gauging one
would expect to find more complicated projectors which are not diagonal in
the indices $i=1,2$ of the supersymmetry parameters $\epsilon ^{i}.$
However, the conditions for the existence of the projectors as well as the
compatibility of the various supersymmetry transformations are very
restrictive. BPS solutions for other types of gauging is currently under
investigation. Other cases with non trivial hypermultiplet dynamics, are the
instanton solutions of the (Euclideanized) $N=2$ supergravity \cite
{sugrainst}. It would be interesting to see whether BPS instanton solutions
exist in the gauged supergravity.

\bigskip

\begin{center}
\textbf{Acknowledgments}.
\end{center}

\medskip\ We are grateful to K. Behrndt for useful discussions and A. Ceresole
for a clarifying correspondence. The research
of M.G. is supported in part by the David and Lucille Packard Foundation.
M.G. is grateful to the Caltech Particle Theory group for hospitality while 
this note  was finalized.
\vskip1truecm

\textbf{Appendix A: The universal hypermultiplet}

\setcounter{equation}{0} \makeatletter
\@addtoreset{equation}{section} \makeatother
\renewcommand{\theequation}{A.\arabic{equation}}

In this appendix we gather some useful formulae for the universal
hypermultiplet and its quaternionic geometry, for more details see \cite
{Ferrara,strominger,lustc}

The universal hypermultiplet moduli space is determined by the K\"{a}hler
potential

\begin{equation}
\phi =-\frac{1}{2}\ln \left( \frac{1}{2}(S+\bar{S}-2|C|^{2})\right).
\end{equation}
Using the coordinates $q^{u}=(S,\bar{S},C,\bar{C})$, the metric components
are
\begin{equation}
g_{S\bar{S}}=\frac{1}{4}e^{4\phi },\text{ \ \ \ }g_{S\bar{C}}=-{{\frac{1}{2}C%
}}e^{4\phi },\text{ \ \ }g_{\bar{S}C}=-{{\frac{1}{2}}}\bar{C}e^{4\phi },%
\text{ \ \ \ }g_{C\bar{C}}=e^{2\phi }+C\bar{C}e^{4\phi }.
\end{equation}
The metric can be given by is given by
\begin{equation}
ds^{2}=u\otimes \bar{u}+v\otimes \bar{v},
\end{equation}
where we have introduced the vielbein forms by \cite{Ferrara, strominger}

\begin{eqnarray}
u &=&e^{\mathcal{\phi }}dC,\quad  \notag \\
v &=&e^{2\mathcal{\phi }}(\frac{dS}{2}-\bar{C}dC),
\end{eqnarray}
satisfying
\begin{equation}
du={{\frac{1}{2}}}u\wedge (v+\bar{v}),\quad dv=v\wedge \bar{v}+u\wedge \bar{u%
},\quad d\mathcal{\phi }=-{{\frac{1}{2}}}(v+\bar{v}).
\end{equation}
One can define the vielbein $f_{i}^{A}$ where $i,A$ are $SU(2)$ indices.
\begin{equation}
f_{1}^{1}=u,\quad f_{1}^{2}=\bar{v},\quad f_{2}^{1}=v,\quad f_{2}^{2}=-\bar{u%
}.  \label{eq:vilbab}
\end{equation}
In components, we have $f_{i}^{A }=(f_{i}^{A})_{u}dq^{u}$ with
\begin{eqnarray}
&&(f_{1}^{1})_{C}=-(f_{2}^{2})_{\bar{C}}=e^{\mathcal{\phi }},  \notag \\
&&(f_{1}^{2})_{\bar{S}}=(f_{2}^{1})_{S}={{\frac{1}{2}}}e^{2\mathcal{\phi }}, 
\notag \\
&&(f_{1}^{2})_{\bar{C}}=-e^{2\mathcal{\phi }}C,  \notag \\
&&(f_{2}^{1})_{C}=-e^{2\mathcal{\phi }}\bar{C}.
\end{eqnarray}
We also have
\begin{eqnarray}
df_{1}^{1} &=&\frac{1}{2}u\wedge (v+\bar{v}),  \notag \\
df_{1}^{2} &=&-(v\wedge \bar{v}+u\wedge \bar{u}),  \notag \\
df_{2}^{2} &=&-{{\frac{1}{2}}}\bar{u}\wedge (v+\bar{v}),  \notag \\
df_{2}^{1} &=&(v\wedge \bar{v}+u\wedge \bar{u}).\text{ \ \ \ \ \ \ }
\end{eqnarray}
Using the Cartan structure equation, one can then find that the $SU(2)=Sp(1)$
connection appearing in the gravitino supersymmetry transformations is given
by

\begin{equation}
p=\left( 
\begin{array}{cc}
\frac{1}{4}(v-\bar{v}) & -u \\ 
\bar{u} & \frac{1}{4}(\bar{v}-v)
\end{array}
\right).
\end{equation}
In components, the connections are
\begin{eqnarray}
p_{S} &=&-\text{\ }p_{\bar{S}}={{\frac{1}{8}}}\left( 
\begin{array}{cc}
e^{2\mathcal{\phi }} & 0 \\ 
0 & -e^{2\mathcal{\phi }}
\end{array}
\right) ,  \notag \\
p_{C} &=&\frac{1}{4}\left( 
\begin{array}{cc}
-e^{2\mathcal{\phi }}\bar{C} & -4e^{\mathcal{\phi }} \\ 
0 & e^{2\mathcal{\phi }}\bar{C}
\end{array}
\right) ,  \notag \\
p_{\bar{C}} &=&\frac{1}{4}\left( 
\begin{array}{cc}
e^{2\mathcal{\phi }}C & 0 \\ 
4e^{\phi } & -e^{2\mathcal{\phi }}C
\end{array}
\right) .
\end{eqnarray}
The covariantly constant $SU(2)$ triplet of K\"{a}hler is given by 
\begin{equation}
K_{i}^{j}=\left( 
\begin{array}{cc}
{-{\frac{1}{2}}}(u\wedge \bar{u}-v\wedge \bar{v}) & -u\wedge \bar{v} \\ 
\bar{u}\wedge v & {{\frac{1}{2}}}(u\wedge \bar{u}-v\wedge \bar{v})
\end{array}
\right)   \label{eq:kftrip}
\end{equation}
or in components
\begin{eqnarray}
K_{S\bar{S}} &=&\frac{1}{8}\left( 
\begin{array}{cc}
e^{4\mathcal{\phi }} & 0 \\ 
0 & -e^{4\mathcal{\phi }}
\end{array}
\right),  \notag \\
K_{C\bar{C}} &=&\frac{1}{2}\left( 
\begin{array}{cc}
-e^{2\mathcal{\phi }}(1-e^{2\mathcal{\phi }}C\bar{C}) & 2e^{3\mathcal{\phi }%
}C \\ 
2e^{3\mathcal{\phi }}\bar{C} & e^{2\mathcal{\phi }}(1-e^{2\mathcal{\phi }}C%
\bar{C})
\end{array}
\right),  \notag \\
K_{S\bar{C}} &=&\frac{1}{4}\left( 
\begin{array}{cc}
-e^{4\mathcal{\phi }}C & 0 \\ 
-2e^{3\mathcal{\phi }} & e^{4\mathcal{\phi }}C
\end{array}
\right),  \notag \\
\text{\ \ }K_{\bar{S}C} &=&\frac{1}{4}\left( 
\begin{array}{cc}
e^{4\mathcal{\phi }}\bar{C} & 2e^{3\mathcal{\phi }} \\ 
0 & -e^{4\mathcal{\phi }}\bar{C}
\end{array}
\right)
\end{eqnarray}
and it satisfies 
\begin{equation}
dp+p\wedge p =K.
\end{equation}
Which is equivalent to the statement that the geometry of the universal
hypermultiplet scalars is quaternionic.

\bigskip

\bigskip

\textbf{Appendix B: Gauging the shift symmetry}

\setcounter{equation}{0} \makeatletter
\@addtoreset{equation}{section} \makeatother
\renewcommand{\theequation}{B.\arabic{equation}}

Given a Killing vector of the hypermultiplet manifold $k^{u},$ the $SU(2)$
prepotential is determined by the equation

\begin{equation}
k^{u}K_{uv}=\partial _{v}P+[p_{v},P].  \label{eq:prepk}
\end{equation}

The Killing vector associated with shifts of the axion $a$ is given by

\begin{equation}
k^{u}=(i,-i,0,0)  \label{eq:axionkill}
\end{equation}

The prepotential related by (\ref{eq:prepk}) to this Killing vector is given
by

\begin{equation}
P_{i}^{j}=\left( 
\begin{array}{cc}
-i{\textstyle{\frac{1}{4}}}e^{2\phi } & 0 \\ 
0 & i{\textstyle{\frac{1}{4}}}e^{2\phi }
\end{array}
\right)   \label{eq:killprep}
\end{equation}
The potential produced by the gauging is given by \cite{ceresole} 
\begin{equation}
V=-2g^{2}P_{ij}P^{ij}+2g^{2}N_{iA}N^{iA},
\end{equation}
where $N_{iA}$ is defined by 
\begin{equation}
N^{iA}={\frac{\sqrt{6}}{4}}k^{u}f_{u}^{Ai}.
\end{equation}
Using the definition of $P_{ij}$ and the fact that $f_{u}^{Ai}f_{Aiv}=g_{uv}$
we find 
\begin{equation}
V={\frac{g^{2}}{8}}e^{4\phi }.
\end{equation}

\vfill\eject

\end{document}